\documentclass{elsart}

\journal{Physics Letters, Section A}
\date{21 March 2001}
\def\pmb#1{\leavevmode\setbox0=\hbox{#1}%
\kern-.02em\copy0\kern-\wd0
\kern.04em\copy0\kern-\wd0
\kern-.02em\raise.04em\box0 }

\def\biglb#1{\leavevmode\hbox{\pmb{$\bigl#1$}}}

\def\bigrb#1{\leavevmode\hbox{\pmb{$\bigr#1$}}}

\newcommand{\quph}[1]{quant-ph/#1}
\newcommand{\bU}{\barr{U}}
\newcommand{\barr}[1]{\overline{#1}}
\newcommand{\lkl}{\left(}\newcommand{\rkl}{\right)}

\newcommand{\phpr}{^{\phantom{\prime}}}
\newcommand{\phbig}{\vphantom{\big|}}

\newcommand{\spinhalf}{\mbox{spin-$\half$}}
\newcommand{\spinone}{\mbox{spin-$1$}}
\newcommand{\pdelta}{\delta^{(p)}}
\newcommand{\bra}[1]{\left\langle{#1}\right|}
\newcommand{\bbra}[1]{\bigl\langle{#1}\bigr|}
\newcommand{\ket}[1]{\left|{#1}\right\rangle}

\newcommand{\olket}[1]{|{#1}\rangle}
\newcommand{\braket}[2]{\langle{#1}|{#2}\rangle}

\newcommand{\tr}[1]{{\mathrm{tr}}\left\{ #1 \right\}}
\newcommand{\Exp}[1]{{\mathrm{e}}^{\mbox{\footnotesize$\textstyle{#1}$}}}
\makeatletter
\newcommand{\Text}[1]{%
{\mathchoice{\text@D{#1}}{\text@T{#1}}{\text@F{#1}}{\text@S{#1}}}}
\newcommand{\text@D}[1]{\mbox{\normalsize\textrm{#1}}}
\newcommand{\text@T}[1]{\mbox{\normalsize\textrm{#1}}}
\newcommand{\text@F}[1]{\mbox{\footnotesize\textrm{#1}}}
\newcommand{\text@S}[1]{\mbox{\scriptsize\textrm{#1}}}
\makeatother
\begin{document}

\begin{frontmatter}
\title{The mean king's problem:\\ Prime degrees of freedom}
\author[ATI,MPQ]{Berthold-Georg Englert,}
\author[TA,SC]{Yakir Aharonov}
\address[ATI]{Atominstitut, Technische Universit\"at Wien\\
Stadionallee 2, 1020 Wien, Austria}
\address[MPQ]{Max-Planck-Institut f\"ur Quantenoptik\\
Hans-Kopfermann-Strasse 1, 85748 Garching, Germany}
\address[TA]{School of Physics and Astronomy, Tel-Aviv University\\
Tel-Aviv 69978, Israel}
\address[SC]{Physics Department, University of South Carolina\\
Columbia, SC 29208, USA}
\begin{abstract}
We show how one can ascertain the values of a complete set of 
mutually complementary observables of a prime degree of freedom.
\end{abstract}
\begin{keyword}
Quantum Kinematics, Complementary Observables
\end{keyword}
\end{frontmatter}

\section{Introduction}\label{sec:Intro}
In 1987, one of us (YA) co-authored a paper \cite{VAA} 
with the somewhat provocative title
``How to ascertain the values of $\sigma_x$, $\sigma_y$, and $\sigma_z$
of a spin-$\frac{1}{2}$ particle.''
It reports the solution of what later became known as 
\emph{The King's Problem}:
A mean king challenges a physicist, who got stranded on the remote 
island ruled by the king, to prepare a  \spinhalf\ atom in any state of her
choosing and to perform a control measurement of her liking.
Between her preparation and her measurement, the king's men determine the
value of either $\sigma_x$, or $\sigma_y$, or $\sigma_z$.
Only after she completed her control measurement, the physicist is told which
spin component has been measured, and she must then state the result of that
intermediate measurement correctly.
How does she do it?

This thought experiment has not been realized as yet.
But recently an optical analog has been formulated
\cite{EKW}, and experimental data should be at hand shortly.
Somewhat unexpectedly, and rather rewardingly, 
the photon version of the king's problem suggested a new scheme for 
quantum crypto\-graphy \cite{BEKW}.

Also very recently, we reported a generalization of the king's problem
\cite{YA+BGE:MK1} where, instead of the traditional \spinhalf\ atom, 
a \spinone\ atom is used.
This generalization required answers to two questions:
What are the appropriate \spinone\ analogs of the \spinhalf\
observables $\sigma_x$, $\sigma_y$, $\sigma_z$?
And, how does the physicist rise to the challenge now?

In the present paper we deal with the further generalization to arbitrary 
prime degrees of freedom, where measurements can have at most  
$p$ different outcomes, $p$ being any prime number.
Of course, the situations of Refs.\ \cite{VAA} and \cite{YA+BGE:MK1}, 
\spinhalf\ ($p=2$) and \spinone\ ($p=3$), respectively, are
particular realizations of the prime case.
We believe that this extension of the idea of Ref.\ \cite{VAA} teaches us a
potentially important lesson about the mathematical structure of quantum
kinematics. 

In Sec.\ \ref{sec:CSMCO} we answer the general-prime version of the first 
question asked above. 
The analogs of the three \spinhalf\ observables are identified
as complete sets of mutually complementary observables.
Then the answer to the second question is given in Sec.\ \ref{sec:SavesNeck};
it employs essentially the same strategy that works in the cases of \spinhalf\
and \spinone, so that we have a genuine generalization indeed.
We leave it as a moot point whether generalizations to non-prime degrees of
freedom are possible, or if there are analogs of the variants of the
\spinhalf\ problem that were found by Ben-Menahem \cite{B-M} and Mermin
\cite{Merm}.  
Also, we do not address the intriguing question of whether the geometrical
reasoning that works so well in the \spinhalf\ case \cite{YA+LV} 
lends itself to generalizations for \spinone\ or richer degrees of freedom. 

\section{Pairwise complementary observables}\label{sec:CSMCO}
The three \spinhalf\ observables $\sigma_x$, $\sigma_y$, $\sigma_z$ are
\emph{complete} in the sense that the probabilities for finding their
eigenvalues as the results of measurements specify uniquely the statistical
operator that characterizes the \spinhalf\ degree of freedom of the ensemble
under consideration.
They are not overcomplete because this unique specification is not ensured if
one of the spin components is left out.

In addition to being complete, the observables $\sigma_x$, $\sigma_y$,
$\sigma_z$ are also pairwise \emph{complementary}, which is to say that in a
state where one of them has a definite value, all measurement results for the
other ones are equally probable.
For example, if $\sigma_x=1$ specifies the ensemble, say, then the results of
$\sigma_y$ measurements are utterly unpredictable: $+1$ and $-1$ are found
with equal frequency; and the same is true for $\sigma_z$ measurements.

What is essential here are not the eigenvalues of  $\sigma_x$, $\sigma_y$,
$\sigma_z$, but their sets of eigenstates.
In technical terms, the fact that the transition probabilities
\begin{equation}\label{eq:2.0}
\begin{array}[b]{rcl}
\biglb|\braket{\sigma_x=\pm1}{\sigma_y=\pm1}\bigrb|^2&=&\half\,,\\
\biglb|\braket{\sigma_y=\pm1}{\sigma_z=\pm1}\bigrb|^2&=&\half\,,\\
\biglb|\braket{\sigma_z=\pm1}{\sigma_x=\pm1}\bigrb|^2&=&\half\,,  
\end{array}
\end{equation}
do not depend on the quantum numbers $\pm1$, is the statement of the pairwise 
complementary nature of $\sigma_x$, $\sigma_y$, and $\sigma_z$.
Their algebraic completeness is then an immediate consequence of the
insight that a \spinhalf\ degree of freedom can have at most three
mutually complementary observables.

More generally, there can be no more than $p+1$ such observables for a 
degree of freedom with a $p$-di\-men\-sional space of state vectors \cite{WooFie}.
Following Weyl \cite{Weyl1,Weyl2} and Schwinger 
\cite{JS1,JS2,JS3}, we'll find it convenient to deal with
unitary operators, rather than the hermitian operators to which they would be
closely related.
Thus the $p+1$ observables $U_0,U_1,\dots,U_p$ are unitary and of period $p$,
\begin{equation}
U_m^p=1\,,\qquad
U_m^r\neq1\enskip\Text{if}\enskip r=1,2,\dots,p-1\,,
  \label{eq:2.1}
\end{equation}
for $m=0,1,\dots,p$.
The eigenvalues of each $U_m$ are powers of 
\begin{equation}
  \label{eq:2.2}
  q\equiv\Exp{2\pi i/p}\,,
\end{equation}
the basic $p$-th root of unity, and we denote by $\ket{m_k}$ the $k$-th
eigenstate of $U_m$, so that
\begin{equation}
  \label{eq:2.3}
  U_m\ket{m_k}=\ket{m_k}q^k
\end{equation}
for $m=0,1,\dots,p$ and $k=1,2,\dots,p$.

Both the orthonormality of the $\ket{m_k}$'s for each $m$ and the mutual
complementarity for different $m$'s are summarized in
\begin{equation}\label{eq:2.4}
\begin{array}[b]{rcl}
  \biglb|\braket{m\phpr_k}{m'_{k'}}\bigrb|^2&=&
\delta_{mm'}\delta_{kk'}+\frac{1}{p}\bigl(1-\delta_{mm'}\bigr)
\\ &=&
\left\{\begin{array}{c@{\quad\Text{if}\quad}l}
\delta_{kk'}& m=m'\,, \\[1ex] p^{-1} & m\neq m'\,,
\end{array}\right.
\end{array}
\end{equation}
for $m,m'=0,1,\dots,p$ and $k,k'=1,2,\dots,p$.
With
\begin{equation}
  \label{eq:2.5}
  U_m=\sum_{k=1}^p\ket{m_k}q^k\bra{m_k}
\end{equation}
this implies
\begin{equation}
  \label{eq:2.6}
  p^{-1}\tr{U_m^r U_{m'}^s}
=\delta_{mm'}\pdelta_{r,-s}
+\bigl(1-\delta_{mm'}\bigr)\pdelta_{r,0}\pdelta_{s,0}\,,
\end{equation}
where $m,m'=0,1,\dots,p$ and $r,s=0,\pm1,\pm2,\dots$, and
\begin{equation}
  \label{eq:2.7}
  \pdelta_{rs}\equiv
\left\{\begin{array}{l@{\enskip}l}
1&\Text{if}\enskip q^r=q^s\\[1ex]
0&\Text{otherwise}
\end{array}\right\}=\frac{1}{p}\sum_{k=1}^pq^{(r-s)k}
\end{equation}
is the appropriate $p$-periodic version of Kronecker's delta symbol.
The reverse is also true: (\ref{eq:2.6}) implies (\ref{eq:2.4}), as can be
shown with the aid of
\begin{equation}
  \label{eq:2.8}
  \ket{m_k}\bra{m_k}=\frac{1}{p}\sum_{r=1}^p\left(q^{-k}U_m\right)^r\,.
\end{equation}
Thus, given a set of $p+1$ unitary operators of period $p$, we
can verify the defining property (\ref{eq:2.4}) of their pairwise 
complementarity by demonstrating that (\ref{eq:2.6}) holds.

Repeated measurements of the observables $U_m$ 
(on identically prepared  systems) eventually determine the 
probabilities $w^{(m)}_k$ for finding their eigenstates $\ket{m_k}$.
As a consequence of their mutual complementarity, knowledge of the
probabilities for one $U_m$ contains no information whatsoever about the
probabilities for any other one.
These $(p+1)\times p$ probabilities represent $p^2-1$ parameters in total, 
since 
\begin{equation}
  \label{eq:2.8'}
  \sum_{k=1}^p w_k^{(m)}=1
\end{equation}
for each of the $p+1$ measurements.
The statistical operator that characterizes the ensemble of identically
prepared systems,
\begin{equation}
  \label{eq:2.9}
\rho=\sum_{m=0}^p\sum_{k=1}^p\ket{m_k}
     \lkl w^{(m)}_k-\frac{1}{p+1}\rkl\bra{m_k}\,,
\end{equation}
is therefore uniquely determined by the probabilities 
${w^{(m)}_k=\bra{m_k}\rho\ket{m_k}}$.
Indeed, the $U_m$'s constitute a complete set of pairwise complementary
observables for the prime degree of freedom under consideration.

Actually, the prime nature of $p$ has not been significant so far, 
but it is for the explicit construction of the set $U_0,U_1,\dots,U_p$ 
that we turn to now. 
We pick an arbitrary period-$p$ unitary operator for $U_0$.
The unitary operator that permutes the eigenvectors of $U_0$ cyclically is
used for $U_p$.
Its eigenvectors in turn are cyclically permuted by $U_0$, 
so that $U_0$ and $U_p$ are jointly characterized by
\begin{equation}\label{eq:2.10a}
\bra{0_k}U_p=\bra{0_{k+1}}\,,\qquad U_0\ket{p_k}=\ket{p_{k+1}}
\end{equation}
for $k=1,2,\dots,p-1$ and, to complete the cycle,
\begin{equation}\label{eq:2.10b}
\bra{0_p}U_p=\bra{0_1}\,,\quad U_0\olket{p_p}=\ket{p_1}\;.
\end{equation}
The fundamental Weyl commutation relation
\begin{equation}
  \label{eq:2.11}
  U_0U_p=q^{-1}U_pU_0
\end{equation}
is an immediate consequence of this reciprocal definition of $U_0$ and $U_p$.
The other $U_m$'s are chosen as
\begin{equation}
  \label{eq:2.12}
  U_m=U_0^mU_p\,.
\end{equation}
Since $p$ is a prime --- what follows is not true for composite numbers;
try $p=6$, for instance, to see what goes wrong --- 
the powers of the $U_m$'s that appear in (\ref{eq:2.8}) comprise all products
of powers of $U_0$ and $U_p$, and since the unitary operators
\begin{equation}
  \label{eq:2.13}
  U_0^rU_p^s\enskip\Text{with}\enskip r,s=1,2,\dots,p\,,
\end{equation}
which are $p^2$ in number, are a basis in the $p^2$ dimensional operator 
algebra \cite{Weyl1,Weyl2,JS1,JS2,JS3}, the $p^2-1$ unitary operators
\begin{equation}
  \label{eq:2.14}
  U_m^r\enskip\Text{with}\enskip r=1,2,\dots,p-1\,,
\end{equation}
supplemented by $1=U_0^p=U_1^p=\cdots=U_p^p$ are also such an operator basis.
As it should be, these bases are complete, but not overcomplete; 
none of the basis operators is superfluous.

As a consequence of (\ref{eq:2.10a}) and (\ref{eq:2.10b}) 
all operators in (\ref{eq:2.13}) are
traceless with the sole exception of the identity operator that obtains for 
$r=s=p$. It is then a matter of inspection to verify that the $U_m$'s
thus constructed obey (\ref{eq:2.6}) and are, therefore, a set of pairwise
complementary observables, indeed.  
From the point of view of the information-theoretical approach to quantum
mechanics that is being developed by Brukner and Zeilinger \cite{CB+AZ},
the $U_m$'s form a complete set of mutually complementary propositions.

\section{The mean king's problem generalized}\label{sec:SavesNeck}
In the generalized version of \emph{The King's Problem} then, 
either one of the observables $U_0,U_1,\dots,U_p$ 
is measured by the mean king's men, 
on a $p$-system suitably prepared by the physicist.
Without knowing which measurement was done actually, the physicist
performs a subsequent measurement of her own, and --- after then being
told which $U_m$ was measured by the king's men --- she has to state 
correctly what they found: $\ket{m_1}$, or $\ket{m_2}$, \dots, or $\ket{m_p}$.

The physicist solves the problem by first preparing a state $\ket{\Psi_0}$ in
which the given $p$-system, the \emph{object}, is entangled with an auxiliary 
$p$-system, the \emph{ancilla}, whose operators and states are barred for 
distinction.
For the ancilla, there are analogs $\bU_0$ and $\bU_p$ of the 
fundamental Weyl operators $U_0$ and $U_p$ that we have for the object.
It is advantageous, however, to interchange the roles of $\bU_0$ and $\bU_p$
in their reciprocal definition. So, rather than just copying the object
relations (\ref{eq:2.10a}) and (\ref{eq:2.10b}), we write for the ancilla
\begin{equation}
    \label{eq:3.0a}
    \bU_p\ket{\barr{0}_k}=\ket{\barr{0}_{k+1}}\,,\quad
    \bra{\barr{p}_k}\bU_0=\bra{\barr{p}_{k+1}}
\end{equation}
for $k=1,2,\dots,p-1$ and
\begin{equation}
    \label{eq:3.0b}
   \bU_p\ket{\barr{0}_p}=\ket{\barr{0}_1}\,,\quad
    \bbra{\barr{p}_p}\bU_0=\bra{\barr{p}_1}\,,
\end{equation}
and the corresponding analog of (\ref{eq:2.12}) is
\begin{equation}
  \label{eq:3.1}
  \bU_m=\bU_p\bU_0^{\,m}
\end{equation}
for $m=1,\dots,p-1$.
Then the transition amplitudes $\braket{0_j}{m_k}$ and 
$\braket{\barr{0}_j}{\barr{m}_k}$
between the eigenstates of $U_0$ and $U_m$ and between those of 
$\bU_0$ and $\bU_m$, respectively, obey recurrence relations,
\begin{equation}
  \label{eq:3.2}
  \frac{\braket{0_{j+1}}{m_k}}{\braket{0_j}{m_k}}=q^{-jm+k}\,,\quad
\frac{\braket{\barr{0}_{j+1}}{\barr{m}_k}}{\braket{\barr{0}_j}{\barr{m}_k}}
=q^{jm-k}
\end{equation}
(for $m\neq0$, of course),
which allow and invite to choose the phase conventions such that
\begin{equation}
  \label{eq:3.3}
  \braket{0_j}{m_k}=\braket{\barr{m}_k}{\barr{0}_j}\,.
\end{equation}
We note in passing that the $\ket{m_k}'s$, or the $\ket{\barr{m}_k}$'s, 
are essentially identical with the states found by Wootters and Fields 
\cite{WooFie} if one opts for the solutions
\begin{equation}
  \label{eq:3.2'}
  \braket{0_j}{m_k}=p^{-1/2}q^{jk-j(j-1)m/2}=\braket{0_j}{\barr{m}_k}^*
\end{equation}
of the recursions (\ref{eq:3.2}).

Joint states in which the object is in 
$\ket{m'_{k'}}$ and the ancilla in $\ket{\barr{m}_k}$ are
denoted by $\ket{m'_{k'}\barr{m}\phpr_{k}}$.
Then
\begin{equation}
  \ket{\Psi_0}=p^{-1/2}\sum_{k=1}^p\ket{m_k\barr{m}_k}
\label{eq:3.4}
\end{equation}
is the entangled object-ancilla state that the physicist prepares.
Thanks to the phase conventions (\ref{eq:3.3}), the $m$ dependence  
is only apparent. 
For either value of $m=0,1,\dots,p$ we get the same $\ket{\Psi_0}$.

If the king's men then measure the object observable $U_m$ and find the
eigenvalue $q^k$, the resulting object-ancilla state is $\ket{m_k\barr{m}_k}$.
After their measurement, there are thus all together $p+1$ sets 
(labeled by $m$) of $p$ possible object-ancilla states each.
These ${(p+1)\times p}$ states cannot be linearly independent because the state
space is only $p^2$-dimensional.
Indeed, each of the $p$-dimensional subspaces spanned by the $p+1$ sets
contains $\ket{\Psi_0}$ by construction. 
In addition, there are ${(p+1)\times(p-1)=p^2-1}$ other states, and we now
proceed to show that they \emph{are} linearly independent.

Consider $m\neq m'$ and any pair of values for $k$ and $k'$.
Then
\begin{equation}
  \label{eq:3.5}
  \braket{m_k\barr{m}_k}{m'_{k'}\barr{m'}_{k'}}=p^{-1}
\end{equation}
as a consequence of (\ref{eq:2.4}) and (\ref{eq:3.3}), and the definition
(\ref{eq:3.4}) of $\ket{\Psi_0}$ implies
\begin{equation}
  \label{eq:3.6}
  \braket{\Psi_0}{m_k\barr{m}_k}=p^{-1/2}=
  \braket{\Psi_0}{m'_{k'}\barr{m'}_{k'}}\,.
\end{equation}
The two vectors
\begin{equation}
  \label{eq:3.7}
  \ket{m_k\barr{m}_k}-p^{-1/2}\ket{\Psi_0}\,,\quad
  \ket{m'_{k'}\barr{m'}_{k'}}-p^{-1/2}\ket{\Psi_0}
\end{equation}
are therefore orthogonal to $\ket{\Psi_0}$ and orthogonal to each other.
Accordingly, $\ket{\Psi_0}$ together with the $p^2-1$ vectors 
$\ket{\Psi_0},\ket{\Psi_1},\dots,\olket{\Psi_{p^2-1}}$ that are defined by
\begin{equation}
  \label{eq:3.8}
  \ket{\Psi_{(p-1)m+j}}=p^{-1/2}\sum_{k=1}^p\ket{m_k\barr{m}_k}q^{-jk}
\end{equation}
for $m=0,1,2,\dots,p$ and $j=1,2,\dots,p-1$  constitute an orthonormal basis,
\begin{equation}
  \label{eq:3.9}
  \braket{\Psi_n}{\Psi_{n'}}=\delta_{nn'}
   \enskip\Text{for}\enskip n,n'=0,1,\dots,p^2-1\,,
\end{equation}
in the $p^2$-dimensional object-ancilla state space.

Two $\ket{\Psi_n}$'s that have the same $m$ value in (\ref{eq:3.8}) are
orthogonal by construction.
And if $\ket{\Psi_n}$ and $\ket{\Psi_{n'}}$ belong to different $m$ values,
their orthogonality follows immediately as soon as one replaces 
$\ket{m_k\barr{m}_k}$ in (\ref{eq:3.8}) by the difference of (\ref{eq:3.7}),
which does not alter the value of the sum.

Let us now see how all of this helps the physicist to meet the mean king's
challenge. 
She will be able to state correctly the measurement result found by
the king's men if she can find an object-ancilla observable $P$ with 
eigenstates $\ket{P_1},\ldots,\ket{P_{p^2}}$ such that each $\ket{P_n}$ is
orthogonal to $p-1$ members each of the $p+1$ sets of states that are
potentially the case after the measurement by the king's men.
We characterize the looked-for eigenstates of $P$ by  an ordered set of
numbers $k_0,k_1,\dots,k_p$ that indicate which members they are
\emph{not} orthogonal to, so that
\begin{equation}
  \label{eq:3.10}
  \ket{\bigl[k_0k_1\dots k_p\bigr]}
\end{equation}
has the defining property of being orthogonal to the object-ancilla states that
result when measurements of $U_m$ do \emph{not} give the eigenvalue $q^{k_m}$.

Suppose the physicist finds the state $\ket{\bigl[325\dots7\bigr]}$.
She then knows that if the king's men had measured $U_0$, $U_1$, $U_2$, or
$U_p$, the respective results must have been $q^3$, $q^2$, $q^5$, and $q^7$, 
because she would never find $\ket{\bigl[325\dots7\bigr]}$ for other 
measurement results.

Accordingly, all that is needed to complete the solution of the generalized
mean king's problem is the demonstration that we can have a
complete orthonormal set of object-ancilla states of the kind (\ref{eq:3.10}).
First note that the expansion of $\ket{\bigl[k_0k_1\dots k_p\bigr]}$ in the
$\ket{\Psi_n}$ basis is given by
\begin{equation}
\ket{\bigl[k_0k_1\dots k_p\bigr]}
=\frac{1}{p}\Bigl(\ket{\Psi_0}
+\sum_{m=0}^p\sum_{j=1}^{p-1}q^{jk_m}\ket{\Psi_{(p-1)m+j}}\Bigr)\,.
  \label{eq:3.11}
\end{equation}
Then observe that
\begin{equation}
  \label{eq:3.12}
\bigl\langle\bigl[k\phpr_0k\phpr_1\dots k\phpr_p\bigr]\big|
\bigl[k'_0k'_1\dots k'_p\bigr]\bigr\rangle
=\frac{1}{p}\sum_{m=0}^p\delta_{k\phpr_m,k'_m}-\frac{1}{p}\,,
\end{equation}
so that two such states are orthogonal if $k\phpr_m=k'_m$ for one and only one
$m$ value.
Therefore, a possible choice of basis states for the physicist's final
measurement is given by those $p^2$ states for which
$ k_0,k_1=1,2,\dots,p$ and 
\begin{equation}
  \label{eq:3.13}
   k_m=(m-1)k_0+k_1\enskip(\mathrm{mod}\ p) 
\end{equation}
for $m=2,3,\dots,p$\,.
The prime nature of $p$ is crucial for the otherwise straightforward
demonstration of the orthogonality of two such states that differ in their
values of $k_0$, or $k_1$, or both.

So, the physicist just has to choose her object-ancilla observable $P$ such 
that it distinguishes the states specified in (\ref{eq:3.13}). 
After being told which measurement the king's men performed on the object, 
she can then infer their measurement result correctly, and with 
certainty, in the manner described above for $\ket{\phbig[325\dots7]}$.

\section*{Acknowledgments}
YA gratefully acknowledges the kind hospitality extended to him by Herbert
Walther during a visit at the MPI f\"ur Quantenoptik in Garching. 
BGE would like to thank Wolfgang Schleich and the University of Ulm for
financial support while part of this work was done,
and he expresses his sincere gratitude for the hospitable environment 
provided by Gerald Badurek and Helmut Rauch at the Atominstitut and 
for the financial support by the Technical University of Vienna.
YA's research is supported in part by Grant No.\ 471/98 of the Basic Research
Foundation (administered by the Israel Academy of Sciences and Humanities) and
NSF Grant No.\ \mbox{PHY-9971005}.

\end{document}